\documentclass[12pt]{article}
\pdfoutput=1
\usepackage{tikz}
\usetikzlibrary{matrix,arrows,decorations.pathmorphing,decorations.pathreplacing}
\usepackage{jheppub}
\usepackage{amsmath,amssymb,euscript,array,mathrsfs,appendix,ctable,marvosym}
\usepackage{arydshln}
\usepackage{todonotes}
\usepackage{graphicx}
\usepackage[normalem]{ulem}

\input{tcilatex}

\newcommand{\EQ}[1]{\begin{equation}\begin{split} #1
\end{split}\end{equation}}

\title{Integrable Lambda Models And Chern-Simons Theories}
\author{David M. Schmidtt\footnote{david@df.ufscar.br}}

\affiliation{Departamento de F\'\i sica, Universidade Federal de S\~ao Carlos, \\
Caixa Postal 676, CEP 13565-905, S\~ao Carlos-SP, Brazil} 

\abstract{In this note we reveal a connection between the phase space of lambda models on $S^{1}\times \mathbb{R}$ and the phase space of double Chern-Simons theories on $D\times \mathbb{R}$ and explain in the process the origin of the non-ultralocality of the Maillet bracket, which emerges as a boundary algebra. In particular, this means that the (classical) $AdS_{5}\times S^{5}$ lambda model can be understood as a double Chern-Simons theory defined on the Lie superalgebra $\mathfrak{psu}(2,2|4)$ after a proper dependence of the spectral parameter is introduced. This offers a possibility for avoiding the use of the problematic non-ultralocal Poisson algebras that preclude the introduction of lattice regularizations and the application of the QISM to string sigma models. The utility of the equivalence at the quantum level is, however, still to be explored.}

\begin{document}

\maketitle


\section{Introduction}
It is by now widely recognized that integrability plays a fundamental role on the AdS/CFT correspondence and that a way to explore the duality more efficiently is to study its underlying integrable structure in a systematic way. One logical strategy to do so is to implement deformations in a consistent mathematical way and then learn more about the physical system from its response to the deformation. Recently, two different but complementary kinds of deformations defined on the gravity side of the duality have been introduced \cite{eta-def fer,lambda-fer}. Both preserve the integrability of (super)-string sigma models and are currently known as the eta models \cite{Klimcik,eta-def bos,eta-def fer,eta-def fer 2} and the lambda models \cite{Sfetsos,lambda-bos,lambda-fer,hybrid}. The works \cite{eta-def bos,eta-def fer,eta-def fer 2} and \cite{lambda-bos,lambda-fer,hybrid}, respectively, came as generalizations of the original ideas for deforming sigma models introduced by Klim\v{c}\'{i}k in \cite{Klimcik} for the eta models and by Sfetsos in \cite{Sfetsos} for the lambda models. In the particular case of the $AdS_{5} \times S^{5}$ Green-Schwarz (GS) superstring, the main property is that its eta/lambda model realize a quantum group deformation of their parent sigma model S-matrix with a $q$ that is real and a root-of-unity \cite{S1,S2,S3}, respectively. 

Most of the physically interesting integrable field theories (including the ones mentioned above) are of the so-called non-ultralocal type, a property that poses a major obstacle to the use of powerful techniques like the algebraic Bethe ansatz and this is why a great amount effort has been invested along the years in trying to eliminate this ``pathological'' behavior e.g. see \cite{FR,1,2,3,4,5,Alleviating-bos,6} for several different approaches concerning this issue. The most important work dealing successfully with this problem is the 1986 seminal paper by Faddeev and Reshetikhin (FR) \cite{FR}, in which a (rather ad hoc) ultralocalization method for the $SU(2)$ principal chiral model (PCM) was introduced allowing to exactly quantize the theory within the QISM scheme. Unfortunately, the method only seemed to work with this case and not with the more interesting PCM's on any Lie group $G$ or the more general sigma models on symmetric spaces $F/G$. It was only in 2012 where real progress was made by Delduc, Magro and Vicedo \cite{Alleviating-bos}, in which the underlying algebraic mechanism behind the ultralocalization method of FR was discovered, generalized and applied to any PCM and sigma model on (semi)-symmetric spaces\footnote{The $AdS_{5}\times S^{5}$ superstring was considered in \cite{Alleviating-fer,lattice-Poisson}.}. Unfortunately, in the case of sigma models on (semi)-symmetric spaces the non ultralocality is still present albeit in an alleviated way and the introduction of a lattice regularization (at quantum level) for the alleviated theories is still not known because of the non-ultralocality persists\footnote{To the present, it has been only possible to construct a lattice Poisson algebra that is related to the Pohlmeyer reduction of the string sigma models \cite{Alleviating-bos,Alleviating-fer,lattice-Poisson}.}.

One of the main characteristics of the lambda deformation is that it implements the FR mechanism of \cite{Alleviating-bos} directly at the Lagrangian level \cite{quantum-group} and this is the best we can do (to present knowledge) in handling analytically the non ultra-locality of the integrable field theory from a world-sheet theory point of view. This means, in particular, that the problem is still present so apparently nothing seems to be gained by deforming the original theory in this particular way. However, it is the same deformed theory that suggests there is a way out if we give up the world-sheet description.

In this work we offer a new approach to deal with the non ultralocality of all known lambda models, which have recently attracted a lot of attention. The idea is not to tackle the problem in 1+1 dimensions, as customary, but rather from a 2+1 dimensional point of view. As we shall see, by changing the dimensionality the problem ceases to exist (for any value of the deformation parameter $\lambda$) and the strategy to do it is to exploit the natural relationship that exist between WZW models and Chern-Simons (CS) theories.  We are also able to introduce the spectral parameter in the 2+1 theory giving it a more prominent role. We expect this approach will provide a novel way to treat the 1+1 integrable field theories that fit within the formulation of lambda models but one of the hopes is to leave open the possibility of generalizing the construction so that more general theories can be treated in a similar way. For a new but different approach to non-ultralocal integrable field theories, see the very recent work \cite{dihedral}. See also \cite{QISM-Tim} for another recent application of the QISM to the lambda model of the PCM.

The lambda models have two important characteristic properties that are analogues of similar relations present on ordinary chiral WZW models. They are summarized
in the following pair of (on-shell) results \cite{quantum-group, part II}%
\begin{equation}
m(z_{\pm })=P\exp \Big[ \pm \frac{2\pi }{k}\int\nolimits_{S^{1}}d\sigma 
\mathscr{J}_{\mp }(\sigma)\Big]\text{ \ \ and \ \ }\mathcal{F}=\Psi (z_{+})\Psi
(z_{-})^{-1}. \label{1 eq.}
\end{equation}

In the first equation, $m(z)$ is the monodromy matrix of the 2d theory, $z_{\pm
}=\lambda ^{\pm 1/2}\in 
\mathbb{R}
$ are two special values of the spectral parameter $z$ and $\mathscr{J}_{\pm
}$ are two currents satisfying the algebra of two mutually commuting
Kac-Moody algebras. This relation have been studied in\footnote{This paper is strongly inspired by the results of \cite{hidden}.} \cite{quantum-group} for
bosonic sigma models and after the use of a KM lattice regularization results in the presence of a quantum group symmetry
with a deformation parameter $q$ that is a root of unity. 

In the second equation, we have that $\mathcal{F}$ is the Lagrangian matrix
field entering the definition of the lambda model action and $\Psi (z)$ is the wave
function that appears as the compatibility equation for the Lax pair
representation of the equations of motion \cite{part II,R-matrices}. A similar decomposition appears
for ordinary chiral WZW models but with the very important difference that for the
lambda models the elements $\Psi (z_{\pm })$ are far from being chiral\footnote{Precisely, this decomposition is used in \cite{part I} to construct the deformed giant magnon solutions of lambda models.}. As it is well known \cite{Witten-Jones,Seiberg,zoo}, conventional WZW models are deeply related to 3d Chern-Simons gauge theories and under this connection, the non ultralocality of the Kac-Moody chiral algebras of the WZW model rises as a boundary effect after the impositions of certain constraints on the phase space. We will see below that this situation persist also for lambda models but with the added advantage that a spectral parameter can be naturally introduced and that this time it is the Maillet bracket \cite{Maillet} that emerges as boundary algebra. Hopefully, this remarkable relation will reveal unexpected connections between integrable string sigma models and gauge theories of the CS type that might assist in the quantization of the former theories.

The paper is organized as follows. In section 1, we introduce the lambda models and emphasize the properties that are important for the topic of the present study. In section 2, we elaborate on the version of the Chern-Simon theory that, after introduction of the spectral parameter, turns out to be equivalent to the lambda models at the classical level. We finish with some remarks concerning our approach and mention on problems to be considered in the near future.

\section{Integrable Lambda Models}
In this section we briefly review the most important aspects of the integrable deformations that are of relevance for the present paper. We will restrict the discussion to the specific example of the lambda model of the Green-Schwarz (GS) superstring on the coset superspace $AdS_{5}\times S^{5}$ but also make contact with similar lambda models when useful for clarifying purposes. 

Consider the Lie superalgebra $\mathfrak{f=psu(}2,2|4%
\mathfrak{)}$ of $F=PSU(2,2,|4)$ and its $%
\mathbb{Z}
_{4}$ decomposition induced by the automorphism $\Phi $%
\begin{equation}
\Phi (\mathfrak{f}^{(m)})=i^{m}\mathfrak{f}^{(m)},\text{ \ \ }\mathfrak{f=}%
\bigoplus\nolimits_{i=0}^{3}\mathfrak{f}^{(i)},\text{ \ \ }[\mathfrak{f}%
^{(m)},\mathfrak{f}^{(n)}]\subset \mathfrak{f}^{(m+n)\func{mod}4},\text{ \ \ 
}  \label{auto}
\end{equation}%
where $m,n=0,1,2,3$. From this decomposition we associate the following twisted loop superagebra%
\begin{equation}
\widehat{\mathfrak{f}}=\bigoplus\nolimits_{n\in 
\mathbb{Z}
}\left( \bigoplus\nolimits_{i=0}^{3}\mathfrak{f}^{(i)}\otimes
z^{4n+i}\right) =\bigoplus\nolimits_{n\in 
\mathbb{Z}
}\widehat{\mathfrak{f}}^{(n)},  \label{loop superalgebra}
\end{equation}
which is required to exhibit the integrable properties of the theory in terms of the spectral parameter $z$. Denote by $G$ the bosonic Lie group associated to $\mathfrak{f}^{(0)}=\mathfrak{su}(2,2)\times \mathfrak{su}(4)$.

The lambda model on the semi-symmetric space $F/G$ is defined by the following action functional\footnote{%
The 1+1 notation used in this paper is: $\sigma^{\pm }=\tau\pm \sigma,$ $\partial
_{\pm }=\frac{1}{2}(\partial _{\tau}\pm \partial _{\sigma}),$ $\eta _{\mu \nu
}=diag(1,-1)$, $\epsilon ^{01}=1$, $\delta _{\sigma\sigma ^{\prime }}$=$\delta(\sigma-\sigma^{\prime})$ and $\delta^{\prime} _{\sigma\sigma ^{\prime }}$=$\partial_{\sigma}\delta(\sigma-\sigma^{\prime})$. We also have that $a_{\pm }=\frac{1}{2}%
(a_{\tau}\pm a_{\sigma}).$} \cite{lambda-fer}
\begin{equation}
S=S_{F/F}(\mathcal{F},A_{\mu })-\frac{k}{\pi }\dint_{\Sigma
}d^{2}\sigma \left\langle A_{+}(\Omega -1)A_{-}\right\rangle ,\text{ \ \ }k\in 
\mathbb{Z}
,  \label{deformed-GS}
\end{equation}%
where $\left\langle \ast ,\ast \right\rangle =STr(\ast ,\ast )$ is the
supertrace in some faithful representation of the Lie superalgebra $\mathfrak{f}$, $\Sigma=S^{1}\times \mathbb{R}$ is the world-sheet manifold parameterized by $(\sigma, \tau)$ and 
$\Omega \equiv \Omega(\lambda)$, where
\begin{equation}
\Omega(z) =P^{(0)}+z P^{(1)}+z ^{-2}P^{(2)}+z^{-1}
P^{(3)}  \label{GS-projector}
\end{equation}
is the omega projector characteristic of the GS superstring. The $P^{(m)}$ are projectors along the graded components $\mathfrak{f}^{(m)}$ of $\mathfrak{f}$.
Above, we have that%
\begin{equation}
S_{F/F}(\mathcal{F},A_{\mu })=S_{WZW}(\mathcal{F})-\frac{k}{\pi }%
\dint_{\Sigma }d^{2}\sigma \left\langle A_{+}\partial _{-}\mathcal{FF}^{-1}-A_{-}%
\mathcal{F}^{-1}\partial _{+}\mathcal{F-}A_{+}\mathcal{F}A_{-}\mathcal{F}%
^{-1}+A_{+}A_{-}\right\rangle ,
\end{equation}%
where $S_{WZW}(\mathcal{F})$ is the usual level $k$ WZW model action. The original GS superstring coupling constant is\footnote{To match with the notation of \cite{part I,part II}, take $\kappa^{2}=4\pi g$.} $\kappa^{2}$ and it is related to $k$ through the relation $\lambda ^{-2}=1+\kappa ^{2}/k$. From \eqref{deformed-GS} we realize that the $\lambda$-deformation can be seen as a continuation of the GS superstring into a topological field theory defined by the gauged $F/F$ WZW model.

The gauge field equations of motion are given by%
\begin{equation}
A_{+}=\left( \Omega ^{T}-D^{T}\right) ^{-1}\mathcal{F}^{-1}\partial _{+}%
\mathcal{F},\text{ \ \ }A_{-}=-\left( \Omega -D\right) ^{-1}\partial _{-}%
\mathcal{FF}^{-1},\text{ \ \ }D=Ad_{\mathcal{F}}.  \label{gauge field eom}
\end{equation}%
After putting them back into the action (\ref{deformed-GS}), a
deformation of the non-Abelian T-dual of the GS superstring with respect to the global left action of the supergroup $F$ is
produced. A dilaton is generated in the process but we will not consider its effects here as we are only concerned with the classical aspects of the theory. 

The $\mathcal{F}$ equations of motion, when combined with \eqref{gauge field eom} can be written in two different by equivalent ways
\begin{equation}
\lbrack \partial _{+}+\mathscr{L}_{+}(z_{\pm }),\partial _{-}+\mathscr{L}%
_{-}(z_{\pm })]=0 , \label{both eom}
\end{equation}
where
\begin{equation}
\mathscr{L}%
_{\pm}(z)=I_{\pm}^{(0)}+zI_{\pm}^{(1)}+z^{\pm2}I_{\pm}^{(2)}+z^{-1}I_{\pm}^{(3)}
\label{Light-cone Lax}
\end{equation}%
is the GS superstring Lax pair that besides satisfy the condition
\begin{equation}
\Phi (\mathscr{L}_{\pm }(z))=\mathscr{L}_{\pm }(iz). \label{auto}
\end{equation}
Then, the lambda model equations of motion follow from zero curvature condition of $\mathscr{L}_{\pm}(z)$. Above, the $I_{\pm}^{(m)}$, are the components of the deformed dual currents defined by
\begin{equation}
I_{+}=\Omega ^{T}(z_{+})A_{+},\text{ \ \ }I_{-}=\Omega^{-1} (z_{-})A_{-},\text{ \ \ } z_{\pm}=\lambda^{\pm 1/2}.
\end{equation}

The flatness of the Lax pair is equivalent to the compatibility condition
\begin{equation}
(\partial _{\mu }+\mathscr{L}_{\mu }(z))\Psi (z)=0, \label{compatibility}
\end{equation}
where $\Psi (z)$ is the so-called wave function. This last equation together with \eqref{gauge field eom} and \eqref{Light-cone Lax} allow to relate (on-shell) the Lagrangian fields of the lambda model to the wave function \cite{part II,R-matrices}. For example,
\begin{equation}
\mathcal{F}=\Psi (z_{+})\Psi (z_{-})^{-1},\text{ \ \ }A_{\pm }=-\partial
_{\pm }\Psi (z_{\pm })\Psi (z_{\pm })^{-1}.
\end{equation}

The spatial component of the Lax pair $\mathscr{L}_{\sigma }(z)\equiv 
\mathscr{L}(z)$ satisfy%
\begin{equation}
\mathscr{L}(z_{\pm})=\mp \frac{2\pi }{k}\mathscr{J}_{\mp }, \label{KM at poles}
\end{equation}%
where the currents $\mathscr{J}_{\pm }$ obey the relations of two mutually commuting
Kac-Moody algebras\footnote{The Kac-Moody algebras are protected and does not change under the Dirac procedure \cite{lambda-bos} meaning we can use them on the constrained surface defined by \eqref{gauge field eom}.}
\begin{equation}
\{\overset{1}{\mathscr{J}_{\pm }}(\sigma),\overset{2}{\mathscr{J}_{\pm }}(\sigma^{\prime})\}=-[C_{12},\overset{2}{\mathscr{J}_{\pm }}(\sigma^{\prime})]\delta _{\sigma
\sigma ^{\prime }}\mp \frac{k}{2\pi }C_{12}\delta _{\sigma \sigma ^{\prime
}}^{\prime }. \label{lambda KM}
\end{equation}
Equation \eqref{KM at poles} is valid for all lambda models and as a consequence of this the first relation in \eqref{1 eq.} provide conserved Lie-Poisson charges \cite{quantum-group}.
On the constrained surface defined by \eqref{gauge field eom} the KM currents take the form
\begin{equation}
\mathscr{J}_{+}=\frac{k}{2\pi}(\Omega ^{T}A_{+}-A_{-}),\text{ \ \ }\mathscr{J}_{-}=-\frac{k}{2\pi}(A_{+}-\Omega A_{-}) \label{KM on-shell}
\end{equation}
and are used to relate $\mathscr{J}_{\pm}$ with the deformed dual currents $I_{\pm}$. This is a particularly useful relation because it means the current algebra for $I_{\pm}$ follows from the algebra \eqref{lambda KM}.

By adding to the Lax operator arbitrary $z$-dependent terms proportional to the Hamiltonian
constraints (bosonic and fermionic) of the theory and by demanding that the condition \eqref{auto} and the equation \eqref{KM at poles} are still valid, we obtain the Hamiltonian
or extended Lax operator \cite{hybrid,part II,R-matrices}
\begin{equation}
\begin{aligned}
\mathscr{L}^{\prime }(z)=-\frac{2\pi}{k} \frac{(z^{4}-z_{+}^{4})}{(z_{+}^{4}-z_{-}^{4})}&\left\{ \mathscr{J}%
_{+}^{(0)}+\frac{z_{-}^{3}}{z^{3}}\mathscr{J}_{+}^{(1)}+\frac{z_{-}^{2}}{%
z^{2}}\mathscr{J}_{+}^{(2)}+\frac{z_{-}}{z}\mathscr{J}_{+}^{(3)}\right\}\\
&-\frac{2\pi}{k} \frac{(z^{4}-z_{-}^{4})}{(z_{+}^{4}-z_{-}^{4})}\left\{ \mathscr{J}_{-}^{(0)}+\frac{z_{+}^{3}}{z^{3}%
}\mathscr{J}_{-}^{(1)}+\frac{z_{+}^{2}}{z^{2}}\mathscr{J}_{-}^{(2)}+\frac{%
z_{+}}{z}\mathscr{J}_{-}^{(3)}\right\} .  \label{extended Lax hybrid}
\end{aligned}
\end{equation}%
Then, as a consequence of the Kac-Moody algebra structure of the theory \eqref{lambda KM}, the
Hamiltonian Lax operator obeys the Maillet algebra%
\begin{equation}
\{\overset{1}{\mathscr{L}^{\prime}}(\sigma,z),\overset{2}{\mathscr{L}^{\prime}}(\sigma^{\prime},w)\}=-[\mathfrak{r}%
_{12},\overset{1}{\mathscr{L}^{\prime}}(\sigma,z)+\overset{2}{\mathscr{L}^{\prime}}(\sigma^{\prime},w)]\delta
_{\sigma \sigma^{\prime}}+[\mathfrak{s}_{12},\overset{1}{\mathscr{L}^{\prime}}(\sigma,z)-\overset{2}{\mathscr{L%
}^{\prime}}(\sigma^{\prime},w)]\delta _{\sigma \sigma^{\prime}}-2\mathfrak{s}_{12}\delta _{\sigma \sigma^{\prime}}^{\prime },
\label{Maillet-lambda}
\end{equation}%
which reduce to the two mutually commuting Kac-Moody algebras at the
special points $z_{\pm }$. We will deduce this bracket from a Chern-Simons theory
point of view and write down the explicit form of the $\mathfrak{r/s}$ operators below. It is important to mention that both GS and hybrid superstring formulations share the same extended Lax operator \cite{hybrid} but defined in terms of the Lie superalgebras $\mathfrak{psu}(2,2|4)$ and $\mathfrak{psu}(1,1|2)$, respectively.

The last piece of information is related to the imposition of the Virasoro constraints $T_{\pm\pm}\approx 0$, which renders the lambda model a string theory\footnote{The lambda models are also consistent superstring theories at the quantum level, as has been recently shown in \cite{lambda-ads2xs2,lambda-ads3xs3,lambda-ads5xs5} for $AdS_{n}\times S^{n}$, $n=2,3,5$.}. The
stress-tensor components of the action \eqref{deformed-GS} are given by
\begin{equation}
T_{\pm \pm }=-\frac{k}{4\pi }\big\langle (\mathcal{F}^{-1}D_{\pm }\mathcal{F%
})^{2}+2A_{\pm }(\Omega -1)A_{\pm }\big\rangle ,
\end{equation}%
where $D_{\pm }(\ast )=\partial _{\pm }(\ast )+[A_{\pm },\ast ]$.
On the surface defined by the gauge field equations of motion they reduce to
the usual\ quadratic form albeit in terms of the deformed dual currents
\begin{equation}
T_{\pm \pm }=\frac{k}{4\pi }(z_{+}^{4}-z_{-}^{4})\big\langle I_{\pm
}^{(2)}I_{\pm }^{(2)}\big\rangle ,
\end{equation}%
that in terms of the Lax pair become%
\begin{equation}
T_{\pm \pm }=\pm \frac{k}{4\pi }\big\langle \mathscr{L}_{\pm }^{2}(z_{+})-\mathscr{L}_{\pm }^{2}(z_{-})\big\rangle . \label{T's}
\end{equation}%
From this last expression we can extract the Hamiltonian and momentum densities\footnote{ Use $H=T_{++}+T_{--}$ and $P=T_{++}-T_{--}$} 
\EQ{
H &=\frac{k}{4\pi }\big\langle \mathscr{L}_{\tau }(z_{+})\mathscr{%
L}_{\sigma }(z_{+})-\mathscr{L}_{\tau }(z_{-})\mathscr{L}%
_{\sigma }(z_{-})\big\rangle , \\ \label{Hamiltonian lambda}
P &=\frac{k}{8\pi }\big\langle (\mathscr{L}_{\tau }^{2}(z_{+})+%
\mathscr{L}_{\sigma }^{2}(z_{+}))-(\mathscr{L}_{\tau }^{2}(z_{-})+\mathscr{L}_{\sigma }^{2}(z_{-}))\big\rangle .
}

The expression \eqref{T's} is not unique to the GS superstring and could be considered as a starting point. Indeed, if we take for example the Lax pair for the hybrid superstring on $AdS_{2} \times S^{2}$ given by \cite{hybrid} 
\begin{equation}
\mathscr{L}%
_{+}(z)=I_{+}^{(0)}+zI_{+}^{(1)}+z^{2}I_{+}^{(2)}+z^{3}I_{+}^{(3)},\text{ \
\ }\mathscr{L}%
_{-}(z)=I_{-}^{(0)}+z^{-3}I_{-}^{(1)}+z^{-2}I_{-}^{(2)}+z^{-1}I_{-}^{(3)} ,
\label{hybrid Lax}
\end{equation}%
which also satisfy \eqref{auto} and make use of \eqref{T's}, we do recover the known expressions for the stress-tensor
\begin{equation}
T_{\pm \pm }=\frac{k}{4\pi }(z_{+}^{4}-z_{-}^{4})\big\langle I_{\pm
}^{(2)}I_{\pm }^{(2)}+2I_{\pm
}^{(1)}I_{\pm }^{(3)}\big\rangle 
\end{equation}%
but in terms of a different set of deformed dual currents $I_{\pm}$ written down in \cite{hybrid}. This result also applies to the PCM lambda model but with a different set of points $z_{\pm}$ defined in \cite{quantum-group}.

As we saw above, the lambda models are naturally equipped with two decoupled Kac-moody
algebras and the Lagrangian field decompose in a rather similar way as the Lagrangian field in conventional chiral WZW models. This suggest that the known \cite{Witten-Jones,Seiberg,zoo} relation between WZW models and CS theories could be present for lambda models as well and we now proceed to make this connection more precise.

\section{Double Chern-Simons theory}

Consider the following double Chern-Simons action functional defined by\footnote{We do not know if there is a standard name in the literature for this type of action.} 
\begin{equation}
S_{CS}=S_{(+)}+S_{(-)}, \label{double action}
\end{equation}
where 
\begin{equation}
S_{(\pm )}=\pm \frac{k}{4\pi }\int\nolimits_{M}\big\langle \mathcal{A}%
_{(\pm )}\wedge \hat{d} \mathcal{A}_{(\pm )}+\frac{2}{%
3}\mathcal{A}_{(\pm )}{\wedge \mathcal{A}}_{(\pm )}\wedge \mathcal{A}_{(\pm
)}\big\rangle .  \label{CS copies}
\end{equation}%
The $(\pm )$ sub-index is just a label whose significance will emerge later on, $M
$ is a 3-dimensional manifold and $\mathcal{A}_{(\pm )}$ are two different
3-dimensional gauge fields valued in the Lie superalgebra $\mathfrak{f}$. In what follows we will study
the generic action    
\begin{equation}
S=\frac{\overline{k}}{4\pi }\int\nolimits_{M}\big\langle \mathcal{A\wedge }%
\hat{d} \mathcal{A+}\frac{2}{3}\mathcal{A\wedge A\wedge A}\big\rangle ,%
\text{ \ \ }\overline{k}=\pm k\text{ \ \ for \ \ }(\pm )
\end{equation}
to avoid a duplicated analysis.

In order to define the Hamiltonian theory of our interest we consider the
action on the manifold $M=D\times 
\mathbb{R}
$, where $D$ is a 2-dimensional disc parameterized by $x^{i},$ $i=1,2$ and $%
\mathbb{R}
$ is the time direction parameterized by $\tau $. It is useful to use
radius-angle coordinates $(r,\sigma )$ to describe $D$ as well$.$ In
particular, we use $\sigma $ as a coordinate of $\partial D=S^{1}$ that is identified with the $S^{1}$ entering the definition of the world-sheet $\Sigma=S^{1}\times \mathbb{R}$ of the lambda model action in \eqref{deformed-GS}.

Using the decomposition%
\begin{equation}
\mathcal{A}=d\tau A_{\tau }+A,\text{ \ \ }\hat{d}=d\tau \partial _{\tau
}+d ,
\end{equation}%
we end up with the following action functional
\begin{equation}
S=\frac{\overline{k}}{4\pi }\int\nolimits_{D\times 
\mathbb{R}
}d\tau \left\langle -A\partial _{\tau }A+2A_{\tau }F\right\rangle -\frac{%
\overline{k}}{4\pi }\int\nolimits_{\partial D\times 
\mathbb{R}
}d\tau \left\langle A_{\tau }A\right\rangle , \label{action on DxR}
\end{equation}%
where $F=dA+A^{2}$ is the curvature of the 2-dimensional gauge field $A=A_{i}dx^{i}$ not to be confused with the world-sheet gauge field entering the definition of the action \eqref{deformed-GS}.
Notice that we have omitted the wedge product symbol $\wedge $ in order to simplify
the notation but we can put it back if required. It is also useful to work in terms of differential forms rather than in terms of components.

The Lagrangian is then given by%
\begin{equation}
L=\frac{\overline{k}}{4\pi }\int\nolimits_{D}\left\langle -A\partial _{\tau
}A+2A_{\tau }F\right\rangle -\frac{\overline{k}}{4\pi }\int\nolimits_{%
\partial D}\left\langle A_{\tau }A\right\rangle , \label{double CS Lag}
\end{equation}%
whose arbitrary variation is as follows
\begin{equation}
\delta L=\frac{\overline{k}}{2\pi }\int\nolimits_{D}\left\langle \delta
A_{\tau }F+\delta A(DA_{\tau }-\partial _{\tau }A)\right\rangle +\frac{%
\overline{k}}{4\pi }\int\nolimits_{\partial D}d\sigma \left\langle \delta
A_{\sigma }A_{\tau }-\delta A_{\tau }A_{\sigma }\right\rangle ,
\end{equation}%
where $D(\ast )=d(\ast )+[A,\ast ]$ is a covariant derivative. From this
we find the bulk equations of motion%
\begin{equation}
F=0,\text{ \ \ }\partial _{\tau }A-DA_{\tau }=0,\text{ \ \ on \ \ }D
\label{bulk EOM}
\end{equation}%
stating that the 3-dimensional gauge field $\mathcal{A}$ is flat, as well as
the boundary equations of motion 
\begin{equation}
\left\langle \delta A_{\sigma }A_{\tau }-\delta A_{\tau }A_{\sigma
}\right\rangle =0\text{ \ \ on \ \ }\partial D,  \label{Boundary EOM}
\end{equation}%
which must be solved consistently in order to obtain the field configurations minimizing the action.
A possible useful solution to the boundary equations of motion is to demand that $A_{\tau }=\xi A_{\sigma }$, 
for some constant factor $\xi$ or the more general boundary conditions considered in \cite{Severa}. However, as we shall see, for the lambda models they are automatically satisfied.

The Lagrangian \eqref{double CS Lag} is already written in Hamiltonian form. The Hamiltonian includes a boundary term and it is given
by
\begin{equation}
H=-\frac{\overline{k}}{2\pi }\int\nolimits_{D}\left\langle A_{\tau
}F\right\rangle +\frac{\overline{k}}{4\pi }\int\nolimits_{\partial
D}\left\langle A_{\tau }A\right\rangle.   \label{Hamiltonian}
\end{equation}%
The fundamental Poisson brackets extracted from \eqref{double CS Lag} are found to be\footnote{%
The 2+1 notation used in this paper is: $\epsilon_{12}=1$ and $\delta _{xy}$=$\delta^{(2)}(x-y)$. For the Lie (super)-algebra we define
$\eta _{AB}=\left\langle T_{A},T_{B}\right\rangle ,$ $%
C_{12}$ = $\eta ^{AB}T_{A}\otimes T_{B}$ and $\overset{1}{u}=u\otimes I,$ $\overset{2}{u}%
=I\otimes u$, etc.} 
\begin{eqnarray}
\{\overset{1}{A}_{i}(x),\overset{2}{A}%
_{j}(y)\}=\frac{2\pi }{\overline{k}}\epsilon _{ij}C_{12}\delta _{xy},\text{ \ \ } \label{2d Gauge field PB}
\{\overset{1}{A}_{\tau }(x),\overset{2}{P}_{\tau }(y)\}=C_{12}\delta _{xy} 
\end{eqnarray}%
and for arbitrary functions of $A_{i}$, they generalize to\footnote{%
For arbitrary functions of $A_{\tau },$ the Poisson bracket is obvious and
will not be written.}%
\begin{eqnarray}
\{F(A),G(A)\} &=&\frac{2\pi }{\overline{k}}\epsilon _{ij}\int\nolimits_{D}d^{2}x\frac{%
\delta F(A)}{\delta A_{i}^{A}(x)}\eta ^{AB}\frac{\delta G(A)}{\delta
A_{j}^{B}(x)}.
\end{eqnarray}%
The definition of the functional derivatives $\delta F/\delta A$ to be used in the bracket above is subtle because of the presence of boundaries \cite{Park,Banhados}. To find them, we
start with the variation $\delta F(A)$ and subsequently find a way to write the result as an
integral over the disc $D$ only. For example, for the Hamiltonian we find that
\begin{equation}
\delta H=-\frac{\overline{k}}{2\pi }\int\nolimits_{D}\left\langle \delta
A_{\tau }F+\delta ADA_{\tau }\right\rangle -\frac{\overline{k}}{4\pi }%
\int\nolimits_{\partial D}d\sigma \left\langle \delta A_{\sigma }A_{\tau
}-\delta A_{\tau }A_{\sigma }\right\rangle .
\end{equation}%
Then, to cancel the boundary term we must use the boundary equations of motion (%
\ref{Boundary EOM}) in order to obtain the desired well-behaved result%
\begin{equation}
\delta H=-\frac{\overline{k}}{2\pi }\int\nolimits_{D}\left\langle \delta
A_{\tau }F+\delta ADA_{\tau }\right\rangle .
\end{equation}%

Now we are ready to consider the Dirac procedure. There is a primary
constraint%
\begin{equation}
P_{\tau }\approx 0,
\end{equation}%
whose stability condition leads to a secondary constraint%
\begin{equation}
F\approx 0,  \label{Flatness}
\end{equation}%
which is nothing but the first bulk equation of motion in (\ref{bulk EOM}).
To study the secondary constraints we better introduce the general quantity
\begin{equation}
G_{0}(\eta )=\frac{\overline{k}}{2\pi }\int\nolimits_{D}\left\langle \eta
F\right\rangle 
\end{equation}%
and compute its variation assuming that the test functions $\eta$ are
independent of the phase space variables $\{A_{\tau},A_{i}\}$. We find that%
\begin{equation}
\delta G_{0}(\eta )=\frac{\overline{k}}{2\pi }\int\nolimits_{D}\left\langle
\delta AD\eta \right\rangle +\frac{\overline{k}}{2\pi }\delta
\int\nolimits_{\partial D}\left\langle \eta A\right\rangle .
\end{equation}%
Then, the constraint with a well-defined functional derivative is actually
the shifted one%
\begin{equation}
G(\eta )=G_{0}(\eta )+G_{1}(\eta ),\text{ \ \ }G_{1}(\eta )=-\frac{\overline{%
k}}{2\pi }\int\nolimits_{\partial D}\left\langle \eta A\right\rangle .  \label{generator G}
\end{equation}
Using this we can show that the action of the shifted constraint is a gauge transformation
\begin{equation}
\delta _{\eta }A\equiv \{A,G(\eta )\}=-D\eta \label{gauge transf}
\end{equation}%
and that the second equation of motion in (\ref{bulk EOM}) can be written
as a special gauge transformation%
\begin{equation}
\text{\ }\partial _{\tau }A=\delta _{(-A_{\tau })}A ,
\end{equation}%
because $\delta H=\delta G(-A_{\tau })$. Then, despite of the fact that 
$A_{\tau }$ is a phase space coordinate both quantities turn out to generate the same action. 

The constraint algebra is now
given by the bracket
\EQ{
\{G(\eta ),G(\overline{\eta })\} &=\frac{\overline{k}}{2\pi }%
\int\nolimits_{D}\left\langle D\eta D\overline{\eta }\right\rangle  \\
&=G([\eta ,\overline{\eta }])+\frac{\overline{k}}{2\pi }\int\nolimits_{%
\partial D}\left\langle \eta d\overline{\eta }\right\rangle
}
after some standard manipulations, showing that when the test functions or their
derivatives do not vanish on $\partial D$, the shifted constraints are actually
second class because of the presence of the boundary. On the other hand, the former constraints $G_{0}(\eta)$ are also second class \cite{Park} for the same kind of test functions and this means that no extra gauge-fixing conditions are required allowing to introduce a Dirac bracket for the constraints $F\approx 0$ in a natural way. In this paper we will restrict to this kind of improper \cite{Banhados,Regge-Teitelboim} test/gauge functions only. 

Now we can show that
\begin{equation}
\begin{aligned}
\{G(\eta ),H\} &=-G([\eta ,A_{\tau }])-\frac{\overline{k}}{2\pi }%
\int\nolimits_{\partial D}\left\langle \eta dA_{\tau }\right\rangle  \\
&\approx -\frac{\overline{k}}{2\pi }\int\nolimits_{\partial D}\left\langle
\eta DA_{\tau }\right\rangle .
\end{aligned}
\end{equation}
Using this result we can find the time evolution of the secondary constraints. We obtain
\begin{eqnarray}
\begin{aligned}
\frac{dG(\eta )}{d\tau } &\approx \{G(\eta ),H\}+\int\nolimits_{\partial D}%
\frac{\delta G(\eta )}{\delta \eta ^{A}}\partial _{\tau }\eta ^{A} \\
&\approx \frac{\overline{k}}{2\pi }\int\nolimits_{\partial D}d\sigma
\left\langle \eta F_{\tau \sigma }\right\rangle -\frac{\overline{k}}{2\pi }%
\int\nolimits_{\partial D}d\sigma \partial _{\tau }\left\langle \eta
A_{\sigma }\right\rangle 
\end{aligned}
\end{eqnarray}%
and after pulling $\partial_{\tau}$ outside the integral as $\frac{d}{d\tau}$, we get the final result
\begin{eqnarray}
\begin{aligned}
\frac{dG_{0}(\eta )}{d\tau } &\approx \frac{\overline{k}}{2\pi }\int\nolimits_{\partial D}d\sigma
\left\langle \eta F_{\tau \sigma }\right\rangle , 
\end{aligned}
\end{eqnarray}%
which vanish if we demand that 
\begin{equation}
F_{\tau \sigma }\approx 0\text{ \ \ on \ \ }\partial D. \label{lambda eom}
\end{equation}%
This is the second bulk equation of motion in \eqref{bulk EOM} (or constraint) with $i=\sigma$ extended to $\partial D$. We will come back to this important boundary constraint later on. There are no tertiary constraints.

Following \cite{Park}, we now write down the non-zero Poisson algebra for the quantities $G_{0}, G_{1}, G$ on the constraint surface $F\approx 0$. It is given by\footnote{It is important to mention at this point that $G_{0}$ and $G_{1}$ separately also have well-defined functional variations, as shown in \cite{Park} after a careful treatment of boundary terms.}
\begin{eqnarray}
\begin{aligned}
\{G_{0}(\eta ),G_{0}(\overline{\eta })\} &\approx-\frac{\overline{k}}{2\pi } \int\nolimits_{\partial
D}d\sigma \left\langle \eta ,D_{\sigma }\overline{\eta }\right\rangle ,\\
\{G_{0}(\eta ),G_{1}(\overline{\eta })\} &\approx \frac{\overline{k}}{2\pi } \int\nolimits_{\partial
D}d\sigma \left\langle \eta ,D_{\sigma }\overline{\eta }\right\rangle ,\\
\{G(\eta ),G(\overline{\eta })\} &\approx \frac{\overline{k}}{2\pi } \int\nolimits_{\partial
D}d\sigma \left\langle \eta ,D_{\sigma }\overline{\eta }\right\rangle .\\
\end{aligned} \label{weakly}
\end{eqnarray}%
In order to impose $F\approx 0$ strongly, we
used Dirac brackets. The only non-zero Dirac brackets are easily found to be
\begin{eqnarray}
\begin{aligned}
\{G_{1}(\eta ),G_{1}(\overline{\eta })\}^{\ast } &=\frac{\overline{k}}{2\pi } \int\nolimits_{\partial
D}d\sigma \left\langle \eta ,D_{\sigma }\overline{\eta }\right\rangle ,\\
\{G(\eta ),G(\overline{\eta })\}^{\ast } &=\frac{\overline{k}}{2\pi } \int\nolimits_{\partial
D}d\sigma \left\langle \eta ,D_{\sigma }\overline{\eta }\right\rangle ,\\ 
\end{aligned} \label{bry algebra}
\end{eqnarray}
which are consistent with setting $F=0$ strongly. From this follows that only the boundary contribution $G_{1}(\eta)$ remains. Then, on the constrained surface, \eqref{gauge transf} takes the form \cite{Park}
\begin{equation}
\delta _{\eta }A\equiv \{A,G_{1}(\eta )\}^{*}=-D\eta, \label{gauge transf dirac}
\end{equation}%
which in turn imply that the second set of equations of motion in (\ref{bulk EOM}) can be written as
\begin{equation}
\partial_{\tau }A=\delta _{(-A_{\tau })}A= \{A,H^{*}\}^{*}, 
\end{equation}%
because $\delta H^{*}=\delta G_{1}(-A_{\tau })$. In showing this last result we have used the boundary equations of motion \eqref{Boundary EOM} as required before and the restriction of \eqref{Hamiltonian} to the constrained surface given by
\begin{equation}
H^{*}=\frac{\overline{k}}{4\pi }\int\nolimits_{\partial
D}d\sigma\left\langle A_{\tau }A_{\sigma}\right\rangle.   \label{Hamiltonian dirac}
\end{equation}%
Then, the time evolution on the constraint surface takes the correct form under the Dirac bracket but now in terms of the boundary Hamiltonian. 

To identify the reduced phase space coordinates we replace \eqref{2d Gauge field PB} by its Dirac bracket\footnote{We will not consider phase space functionals depending on $P_{\tau}$.}, but this is equivalent to pulling the symplectic form associated to (the first bracket in) \eqref{2d Gauge field PB} back to the constraint surface \cite{Seiberg,zoo}. Namely,
\begin{equation}
\omega ^{\ast }=\frac{\overline{k}}{4\pi }\int\nolimits_{D}\left\langle
\delta A\wedge \delta A\right\rangle \bigg\rvert_{A=-d\Psi \Psi ^{-1}}=-\frac{%
\overline{k}}{4\pi }\int\nolimits_{\partial D}d\sigma\left\langle \delta A_{\sigma
}\wedge D_{\sigma }^{-1}\delta A_{\sigma }\right\rangle .
\end{equation}
The Poisson bracket that follows\footnote{The symplectic form operator is formally identified as $\hat{\omega}=D_{\sigma}^{-1}$ with an associated Poisson operator given by $\hat{\theta}=D_{\sigma}$ that is responsible for the Kac-Moody algebra structure. \label{15}} from this reduced symplectic form is equivalent to \eqref{bry algebra} and after eliminating the test functions, we reveal the Kac-Moody algebra structure
\begin{equation}
\{\overset{1}{A}_{\sigma }(\sigma ),\overset{2}{A}_{\sigma }(\sigma ^{\prime
})\}^{\ast }=\frac{2\pi }{\overline{k}}\big( [C_{12},\overset{2}{A}_{\sigma }(\sigma ^{\prime })]\delta
_{\sigma \sigma ^{\prime }}+C_{12}\delta _{\sigma \sigma ^{\prime }}^{\prime
}\big)  \label{KM in A}
\end{equation}%
of the associated WZW model on $\partial D$. In this sense we say that the phase space information of the CS theory is now completely stored on its boundary theory. Indeed, the reduced phase space is described by the data $(A_{\sigma}|_{\partial D},H^{*},\{\cdot,\cdot \}^{*})$. The time evolution of $A_{\sigma}$ can be put in Hamiltonian form and it is given precisely by the boundary constraints \eqref{lambda eom}, which as we shall see below are equivalent to the string lambda model equations of motion. What we have presented here is nothing but the Hamiltonian version of the (well-known) relation that exists between the CS and the WZW theories that is found in the literature but in a different guise.

We are now ready to introduce the dependence of the spectral parameter $z$. Not surprisingly, the $(\pm )$
sub-index introduced above make reference to the two special points $%
z_{\pm}=\lambda ^{\pm 1/2}$ in the complex plane parameterized by $z$ and introduced in the last section.
We now make use of the twisted loop superalgebra structure \eqref{loop superalgebra} and consider the problem of finding a $z$-dependent
2-dimensional gauge field $A(z)$ on the disc $D$ satisfying the following two conditions\footnote{Here we are considering only the horizontal fields $A_{(\pm)}$, but it works exactly the same way for $\mathcal{A}_{(\pm)}$ so we can consider $\mathcal{A}(z)$ instead from the beginning and then restrict it to the disc.}.
\begin{equation}
A(z_{\pm})=A_{(\pm )}\text{ \ \ and \ \ }\Phi (A(z))=A(iz). \label{conditions}
\end{equation}%
The answer we will consider here (recall that $A(z)=A_{i}(z)dx^{i}$, $i=1,2$) is given by
\begin{equation}
A(z)=f_{-}(z)\overline{\Omega }(z/z_{+})A_{(+)}-f_{+}(z)\overline{%
\Omega }(z/z_{-})A_{(-)},\text{ \ \ }f_{\pm }(z)=\frac{%
(z^{4}-z_{\pm} ^{4})}{(z_{+}^{4}-z_{-}^{4})},
\label{interpolating current}
\end{equation}%
where%
\begin{equation}
\overline{\Omega }(z)=P^{(0)}+z^{-3}P^{(1)}+z^{-2}P^{(2)}+z^{-1}P^{(3)}
\end{equation}%
is another projector not to be confused with the one defining the GS lambda model \eqref{GS-projector} above. Actually, this same projector appears for both the GS \cite{part II} and the hybrid superstring \cite{hybrid} and leads to the same Maillet bracket when \eqref{extended Lax hybrid} or \eqref{KM in A} is used. Then, both superstring formulations are equivalent at this level of analysis.

Using this $z$-dependent gauge field we can gather both Poisson brackets on the LHS of (\ref{2d Gauge
field PB}) into a single interpolating one
\begin{equation}
\boxed{
\{\overset{1}{A}_{i}(x,z),\overset{2}{A}_{j}(y,w)\}=-2\mathfrak{s}%
_{12}(z,w)\epsilon _{ij}\delta _{xy},}  \label{pre Maillet}
\end{equation}%
which, as we shall see, is the precursor of the Maillet bracket. From here we can appreciate that it is the Chern-Simons Poisson structure ($\{ A_{i},A_{j}\} \approx \epsilon_{ij}$ and not $\{ A_{i},A_{j}\} \approx \delta_{ij}$) the one responsible for the non skew-symmetry of the $R$-matrix entering the Maillet bracket and the very source of its non ultralocality. In this calculation we face exactly the same situation of \cite{hybrid} and find that%
\begin{equation}
\mathfrak{s}_{12}(z,w)=-\frac{1}{z^{4}-w^{4}}\tsum\nolimits_{j=0}^{3}%
\{z^{j}w^{4-j}C_{12}^{(j,4-j)}\varphi _{\lambda
}^{-1}(w)-z^{4-j}w^{j}C_{12}^{(4-j,j)}\varphi _{\lambda }^{-1}(z)\},
\end{equation}%
where%
\begin{equation}
\varphi _{\lambda }(z)=\frac{k}{\pi }\frac{(\lambda ^{-2}-\lambda^{2} )}{%
(z^{-2}-z^{2})^{2}-(\lambda^{-1} -\lambda )^{2}} \label{twisting function}
\end{equation}%
is the lambda deformed twisting function. Notice that the two special points 
$z_{\pm}=\lambda ^{\pm 1/2}$ are poles of $\varphi _{\lambda }(z)$. In retrospective, we realize that our theory \eqref{CS copies} actually consist of two Chern-Simons theories with opposite levels attached to the poles $z_{\pm}$ of \eqref{twisting function} in the complex plane or the Riemann sphere after compactification.

The symmetric operator $\mathfrak{s}_{12}(z,w)$ satisfy%
\begin{equation}
\mathfrak{s}_{12}(z_{\pm},z_{\pm})=\mp \frac{\pi }{k}C_{12}=-%
\frac{\pi }{\overline{k}}C_{12},\text{ \ \ }\mathfrak{s}_{12}(z_{\pm},z_{\mp})=0
\end{equation}%
as required for the Poisson algebra (\ref{pre Maillet}) to reduce to (\ref{2d
Gauge field PB}) at the poles. It also satisfy 
\begin{equation}
\underset{\lambda \rightarrow 0}{\lim }\ \mathfrak{s}_{12}(z,w)=-\frac{\pi }{k}%
C_{12}^{(00)} \label{SG limit}
\end{equation}
but we still do not have a proper interpretation for this limit which corresponds to the ultra-localization limit of the lambda models and that is deeply related to the Pohlmeyer reduction of the $AdS_{5}\times S^{5}$ GS superstring \cite{PR}. As customary, we will refer to the limits $\lambda \rightarrow 0$ and $\lambda \rightarrow 1$ as the sine-Gordon and sigma model limits, respectively.

For arbitrary functions and their differentials
\begin{equation}
F(A)=(F,A)_{\varphi _{\lambda }},%
\text{ \ \ }\underset{t\rightarrow 0}{\lim }\ \frac{d}{dt}F(A+tX)=(dF,X)_{\varphi_{\lambda} },
\end{equation}
the Poisson bracket \eqref{pre Maillet} generalize to
\begin{equation}
\{F(A),G(A)\}=(R(dF),dG)_{\varphi _{\lambda }}+(dF,R(dG))_{\varphi _{\lambda }}, \label{R-bracket}
\end{equation}%
where $R=\pm (\Pi _{\geq 0}-\Pi _{<0})$ is the usual AKS $R$-operator defined in terms of the projectors $\Pi$ that act on elements of $\hat{\mathfrak{f}}$. The definitions are as follows: For a $z$-dependent 2-form on $D$ constructed from the pair $X$ and $Y$, we have 
\begin{equation}
(X,Y)_{\varphi _{\lambda }} =\int_{D}\left\langle X(x),Y(x)\right \rangle_{\varphi _{\lambda }} , \label{inner product on D}
\end{equation}%
where
\begin{equation}
\left\langle X(x),Y(x)\right \rangle_{\varphi _{\lambda }}=\oint_{0}\frac{dz}{2\pi iz}%
\varphi _{\lambda }(z)\left\langle X(x,z),Y(x,z)\right\rangle
\end{equation}
is the twisted inner product on the loop superalgebra $\widehat{\mathfrak{f}}$ for fixed $x$. 
For example, the functions%
\begin{equation}
F_{(\pm )}(A)=\int\nolimits_{D}\big\langle A_{(\pm )}\wedge \eta _{(\pm
)}\big\rangle =\int\nolimits_{D}d^{2}x\big\langle A_{i(\pm )}\eta _{j(\pm
)}\big\rangle \epsilon _{ij},
\end{equation}%
where $\eta _{(\pm )}$ are two test 1-forms can be written in terms of \eqref{inner product on D} as 
\begin{equation}
F_{(\pm )}(A)=(\eta _{(\pm )},A)_{\varphi _{\lambda }},
\end{equation}%
with%
\begin{equation}
\eta _{(\pm )}(z)=\varphi^{-1} _{\lambda }(z)z_{\pm }^{4}(z^{-4}+z_{\mp
}^{4})\{\eta _{(\pm )}^{(0)}+zz_{\mp }\eta _{(\pm )}^{(1)}+z^{2}z_{\mp
}^{2}\eta _{(\pm )}^{(2)}+z^{3}z_{\mp }^{3}\eta _{(\pm )}^{(3)}\}.
\end{equation}%
Now, because of the functions $F_{(\pm )}(A)$ are linear in $A_{(\pm )}$, their
differentials are $dF_{(\pm )}(z)=\eta _{(\pm )}(z).$ Using these expressions
in \eqref{R-bracket}, we recover\footnote{The $R$-operator with the minus sign is the one that reproduce the first Poisson bracket expression in \eqref{2d Gauge field PB} explicitly.} the first expression in \eqref{2d Gauge field PB}.

Now we can compute the $z$-dependent
boundary algebra after imposing the constraints (\ref{Flatness}) strongly. We take $i=\sigma$ in \eqref{interpolating current} and use \eqref{KM in A}. Explicitly,
\begin{equation}
\begin{aligned}
A_{\sigma}(z)=\frac{(z^{4}-z_{-}^{4})}{(z_{+}^{4}-z_{-}^{4})}&\left\{ A_{\sigma(+)}^{(0)}+\frac{z_{+}^{3}}{z^{3}}A_{\sigma(+)}^{(1)}+\frac{z_{+}^{2}}{%
z^{2}}A_{\sigma(+)}^{(2)}+\frac{z_{+}}{z}A_{\sigma(+)}^{(3)}\right\}\\
&-\frac{(z^{4}-z_{+}^{4})}{(z_{+}^{4}-z_{-}^{4})}\left\{ A_{\sigma(-)}^{(0)}+\frac{z_{-}^{3}}{z^{3}%
}A_{\sigma(-)}^{(1)}+\frac{z_{-}^{2}}{z^{2}}A_{\sigma(-)}^{(2)}+\frac{%
z_{-}}{z}A_{\sigma(-)}^{(3)}\right\} .  \label{extended gauge field}
\end{aligned}
\end{equation}%
As a consequence of the Kac-Moody algebra structure \eqref{KM in A} and once more following \cite{hybrid}, we obtain the Maillet bracket
\begin{equation}
\{\overset{1}{A_{\sigma}}(\sigma ,z),\overset{2}{A_{\sigma}}(\sigma
^{\prime },w)\}^{\ast }=-[\mathfrak{r}_{12},\overset{1}{A_{\sigma}}(\sigma
,z)+\overset{2}{A_{\sigma}}(\sigma ^{\prime },w)]\delta _{\sigma \sigma
^{\prime }}+[\mathfrak{s}_{12},\overset{1}{A_{\sigma}}(\sigma ,z)-\overset{2%
}{A_{\sigma}}(\sigma ^{\prime },w)]\delta _{\sigma \sigma ^{\prime }}-2%
\mathfrak{s}_{12}\delta _{\sigma \sigma ^{\prime }}^{\prime },
\label{Maillet}
\end{equation}%
where
\begin{equation}
\mathfrak{r}_{12}(z,w)=\frac{1}{z^{4}-w^{4}}\tsum\nolimits_{j=0}^{3}%
\{z^{j}w^{4-j}C_{12}^{(j,4-j)}\varphi _{\lambda
}^{-1}(w)+z^{4-j}w^{j}C_{12}^{(4-j,j)}\varphi _{\lambda }^{-1}(z)\},
\end{equation}%
is the anti-symmetric part of the $R$-matrix. This is the same algebra obtained with the extended Lax operator \eqref{extended Lax hybrid} and we now identify $A_{\sigma}(z)=\mathscr{L}^{\prime}(z)$. The bracket satisfy the Jacobi identity and reduce to \eqref{KM in A} at the poles. The sine-Gordon limit \eqref{SG limit} applied to \eqref{Maillet} is a continuous version of the alleviating mechanism introduced in \cite{Alleviating-bos} so the non ultralocality of the Maillet algebra is still present for (semi)-symmetric cosets. Then, in order to suppress the $\delta _{\sigma \sigma ^{\prime }} ^{\prime }$ completely for any value of $\lambda$, we must go to a higher dimension.

Alternatively, by setting 
\begin{equation}
\mathfrak{r}_{12}=\frac{1}{2}(R_{12}-R_{12}^{\ast }),\text{ \ \ }\mathfrak{s}%
_{12}=-\frac{1}{2}(R_{12}+R_{12}^{\ast }),
\end{equation}%
we can write\footnote{Up to a global minus sign this is the same Maillet braket constructed in \cite{R-matrices}.}
\begin{equation}
\{\overset{1}{A_{\sigma}}(\sigma ,z),\overset{2}{A_{\sigma}}(\sigma
^{\prime },w)\}^{\ast }=-[R_{12},\overset{1}{A}_{\sigma }(\sigma ,z)]\delta _{\sigma
\sigma ^{\prime }}+[R_{12}^{\ast },\overset{2}{A}_{\sigma }(\sigma ^{\prime
},w)]\delta _{\sigma \sigma ^{\prime }}+(R_{12}+R_{12}^{\ast })\delta
_{\sigma \sigma ^{\prime }}^{\prime },
\end{equation}%
where%
\begin{equation}
R_{12}(z,w)=\frac{2}{z^{4}-w^{4}}\tsum%
\nolimits_{j=0}^{3}z^{j}w^{4-j}C_{12}^{(j,4-j)}\varphi _{\lambda }^{-1}(w),%
\text{ \ \ }R_{12}^{\ast }(z,w)=R_{21}(w,z). \label{R12}
\end{equation}

For arbitrary functions of $A_{\sigma}$, \eqref{Maillet} generalize to
\begin{equation}
\{F,G\}^{*}(A_{\sigma})=-(A_{\sigma},[dF,dG]_{R})_{\varphi
_{\lambda }}+\omega (R(dF),dG)_{\varphi _{\lambda }}+\omega
(dF,R(dG))_{\varphi _{\lambda }} ,  \label{R-Maillet}
\end{equation}%
where $[*,*]_{R}$ is the $R$-bracket on $\widehat{\mathfrak{f}}$ and
\begin{equation}
\omega (X,Y)_{_{\varphi _{\lambda }}} =\int_{\partial D}d\sigma\left\langle X(\sigma),\partial
_{\sigma}Y(\sigma)\right\rangle_{\varphi _{\lambda }} 
\end{equation}
is the co-cycle. The only difference when compared to \eqref{inner product on D} is that the inner product integration is now performed on $\partial D$, the $d\sigma$ is written explicitly and the $X,Y$ are ordinary functions on $\partial D$. Namely,
\begin{equation}
(X,Y)_{\varphi _{\lambda }} =\int_{\partial D} d\sigma \left\langle X(\sigma),Y(\sigma)\right \rangle_{\varphi _{\lambda }} . \label{inner product on boundary}
\end{equation}%

The bracket \eqref{Maillet} can, alternatively, be written in the form
\begin{equation}
\{F,G\}^{*}(A_{\sigma })=(R(dF),\hat{D}_{\sigma}dG)_{\varphi _{\lambda }}+(dF,\hat{D}_{\sigma}R(dG))_{\varphi _{\lambda }},
\end{equation} \
where $\hat{D}_{\sigma}(\ast)=\partial_{\sigma}(\ast)+[A_{\sigma}(z),\ast]$. From this we identify $\hat{\theta}(z)=\hat{D}_{\sigma}\circ R+R^{*}\circ \hat{D}_{\sigma}$, which is the analogue of the $\hat{\theta}$ in \eqref{KM in A} (see footnote \eqref{15}). Following the same steps we realize that \eqref{R-Maillet} is the $z$-dependent extension of the Dirac bracket associated to \eqref{R-bracket} after imposing the constraints $F(z_{\pm})=0$. A comment is in order. Notice that we are referring to \eqref{R-Maillet} as an extension of the Dirac bracket because $F(z)$ (the curvature of $A(z)$) reproduce the correct Hamiltonian constraints only when it reach the poles $z_{\pm}$. In order to find a proper $z$-dependent constraint (if any), we probably would need to lift the action \eqref{double action} to the loop superalgebra $\hat{\mathfrak{f}}$ and run the Dirac procedure again but as we have seen, the introduction of the spectral parameter is rather innocuous and does not introduce new degrees of freedom or fields so no new constraints are expected beyond those attached to the points $z_{\pm}$. However, by an abuse of notation we will keep the $\ast$ on \eqref{R-Maillet}.

Now we are in the position to interpret the boundary equations of motion \eqref{Boundary EOM}. First we
note that the link between the Kac-Moody algebras \eqref{KM in A} and \eqref{lambda KM} is through the
relation
\begin{equation}
A_{\sigma (\pm )}=\mp \frac{2\pi }{k}\mathscr{J}_{\mp }=\mathscr{L}_{\sigma
}(z_{\pm }), \label{sigma}
\end{equation}%
where we have used \eqref{KM at poles} in the last equality. Now, the obvious solution to the
boundary equations of motion is to identify%
\begin{equation}
A_{\tau (\pm )}=\mathscr{L}_{\tau }(z_{\pm }).  \label{tau}
\end{equation}%
To see why, we rewrite \eqref{Boundary EOM} in the form
\begin{equation}
\epsilon ^{\mu \nu }\left\langle \delta \mathscr{L}_{\mu }(z_{+})\mathscr{L}%
_{\nu }(z_{+})-\delta \mathscr{L}_{\mu }(z_{-})\mathscr{L}_{\nu
}(z_{-})\right\rangle =0 \label{bdry}
\end{equation}%
and use the fact that the product $\left\langle \delta \mathscr{L}_{\pm }(z)%
\mathscr{L}_{\mp }(z)\right\rangle $ is independent of the spectral
parameter either for the Green-Schwarz or the hybrid superstring Lax Pairs \eqref{Light-cone Lax}
and \eqref{hybrid Lax}, respectively. The PCM Lax pair also satisfy \eqref{bdry} trivially. Thus, for the lambda models the CS boundary equation of motion \eqref{bdry} is just an identity. The explicit transformation between the components of the CS gauge field on $\partial D \times \mathbb{R}$ and the lambda model gauge field on $\Sigma$ is
\begin{equation}
A_{\sigma (\pm )}=%
\bigg\{ \begin{array}{c}
A_{+}-\Omega A_{-} \\ 
\Omega ^{T}A_{+}-A_{-}%
\end{array}%
,\text{ \ \ }A_{\tau (\pm )}=%
\bigg\{ \begin{array}{c}
A_{+}+\Omega A_{-} \\ 
\Omega ^{T}A_{+}+A_{-}%
\end{array} .
\end{equation}

The relations \eqref{sigma} and \eqref{tau} allow to extract an important piece of information from the boundary constraints \eqref{lambda eom} if we write them in the form
\begin{equation}
\lbrack \partial _{+}+\mathscr{L}_{+}(z_{\pm }),\partial _{-}+\mathscr{L}%
_{-}(z_{\pm })]=0 .
\end{equation}
The first conclusion is that they are precisely the lambda model Euler-Lagrange equations of motion \eqref{both eom} of the action \eqref{deformed-GS}, which also follows from a Lax pair zero curvature condition. The second conclusion is that they imply the conservation of the Poisson-Lie charges $m(z_{\pm})$ in \eqref{1 eq.} (see \cite{quantum-group}). The boundary degrees of freedom of the double CS theory on $\mathfrak{psu}(2,2|4)$ are described by the $AdS_{5}\times S^{5}$ lambda model action \eqref{deformed-GS}. This conclusion also apply to the other models we have considered so far. 

It is important to realize that the identification between \eqref{Maillet} and \eqref{Maillet-lambda} holds for the extended Lax operator. At this point the result is rather generic (recall we only required \eqref{conditions} in the construction) and in order to consider a particular lambda model the omega projector $\Omega$ must be specified as it determines the Hamiltonian constraint structure of the theory under the Dirac procedure. In other words, it determines the decomposition 
\begin{equation}
\mathscr{L}^{\prime}(z)=\mathscr{L}(z)+\text{constraints},
\end{equation}
from where the current algebra of the deformed dual currents $I_{\pm}$ can be computed. It matches the one found by using the direct relation \eqref{KM on-shell}. See \cite{Magro-exchange,Vicedo-exchange} for the the conventional GS superstring and \cite{hybrid} for the lambda model of the hybrid superstring. It is remarkable that the CS theory reproduce the Hamiltonian Lax connection as it has interesting properties, see \cite{Vicedo-exchange} for the specific case GS formulation in relation to the involution properties of the charges extracted from the monodromy matrix. Notice that the twisted loop and the Kac-moody superalgebras combined are the ones responsible for introducing the $R$-matrix with spectral parameter. 

There is a certain amount of freedom in the construction of the
current (\ref{interpolating current}). For instance, one could consider several
copies of the Chern-Simons actions (\ref{CS copies}) in the definition of
the theory involving different WZW levels $k's$, which would alter the first condition in \eqref{conditions} or we can also consider a consistent algebraic modification to the second condition in \eqref{conditions}. Either case, this could lead to more general twisting functions and $\mathfrak{r/s}$ tensors and to novel multiparametric lambda deformations of string sigma models. An example of a consistent modification of the second condition, when $\mathfrak{f}$ is a bosonic Lie algebra, is the PCM. For this case, the loop superalgebra has $\Phi=I$, i.e. no $\mathbb{Z}_{2}$ grading and 
\begin{equation}
\mathscr{L}^{\prime}(z)=\mathscr{L}(z)=f_{-}(z)\mathscr{J}_{+}+f_{+}(z)\mathscr{J}_{-},
\end{equation}
i.e. no first class constraints. The explicit form of the functions $f_{\pm}(z)$ for this case can be found by working out explicitly the Lax pair representation. The $\mathfrak{r/s}$ tensors as well as the twisting function $\varphi_{\lambda}(z)$ and their poles $z_{\pm}$ become those of the lambda deformation of the PCM \cite{quantum-group}. 

On the constrained surface, the complete Hamiltonian \eqref{Hamiltonian dirac} is given by (we drop the $*$)
\begin{equation}
H=\frac{k}{4\pi }\int\nolimits_{\partial
D}d\sigma\big\langle A_{\tau }(z_{+})A_{\sigma}(z_{+})-A_{\tau }(z_{-})A_{\sigma}(z_{-})\big\rangle ,
\end{equation}%
which should be compared with the lambda model Hamiltonian in \eqref{Hamiltonian lambda}. Mimicking \eqref{Hamiltonian lambda}, we define the momentum generator
\begin{equation}
P=\frac{k}{8\pi }\int\nolimits_{\partial
D}d\sigma\big\langle (A_{\tau }^{2}(z_{+})+%
A_{\sigma }^{2}(z_{+}))-(A_{\tau }^{2}(z_{-})+A_{\sigma }^{2}(z_{-}))\big\rangle ,
\end{equation}
which commutes with $H$ under the bracket \eqref{KM in A}. The opposite signs of the levels at $(\pm)$ are important in showing this. We can recover the Virasoro algebra structure of the lambda models if we define the usual $T_{\pm\pm}$ components as in \eqref{T's}. \

Finally, the boundary equations of motion \eqref{Boundary EOM} can be understood as a condition dictating the form of the Lax pair and the boundary constraints \eqref{lambda eom} as a condition dictating the dynamics of the system because of their equivalence to the lambda model Euler-Lagrange equations of motion. This is precisely the content of equation \eqref{compatibility} which summarizes the integrability properties of the system. The flatness condition as well as the analytic properties of $\mathscr{L}_{\pm}(z)$ are known to be preserved by the action of the group of dressing transformations \cite{dressing,BBT,part II} that can be seen now as an infinite dimensional symmetry group of the boundary theory. 

\section{Concluding remarks}\label{conclusions}

The main goal of this paper was to show how the lambda model \eqref{deformed-GS} can be reformulated as a double Chern-Simons theory \eqref{double action} and how the Lax pair representation and the Maillet bracket structure of the lambda model phase space emerge from the CS theory point of view. The strategy is to trade the non-ultralocal Maillet bracket \eqref{Maillet} by the ultralocal CS bracket \eqref{pre Maillet} at the expense of introducing a couple of new constraints \eqref{Flatness}, so the price to pay for the elimination of the problematic $\delta^{\prime}_{\sigma \sigma^{\prime}}$ term is to deal with the quadratic second class constraints $F(z_{\pm})\approx 0$ on the disc. To obtain a string theory the Virasoro constraints $T_{\pm\pm}\approx 0$ as well as the gauge fixing of the kappa symmetry must be taken into account and, fortunately, both can be handled at the same time by means of the light-cone dressing gauge introduced in \cite{part II}, which reduce to picking a particular orbit of the $\lambda$-deformed BMN vacuum solution under the action of the dressing group $\Psi(z)=\chi(z)\Psi_{0}(z)$.

One may wonder what is to be gained in complicating even more the phase space structure of the string lambda models by introducing the constraints $F(z_{\pm})\approx 0$ proper of the CS setting. On the one hand, by doing this we can, not only to suppress completely the non ultralocality of all known $\lambda$-models (i.e. PCM$_{\lambda}$, F/G$_{\lambda}$, GS on $AdS_{5}\times_{\lambda}S^{5}$ and hybrid on $AdS_{2}\times_{\lambda}S^{2}$) but also to do it for any (generic) value of the deformation parameter $\lambda$. On the other hand, the new theory being of the CS type can, in principle, be quantized in a number of ways. In particular, by employing a (disc) lattice algebra regularization \cite{Fock-Rosly,combinatorial I,combinatorial II,Buffenoir-1,Buffenoir-2} or by a path integral approach \cite{Witten-Jones,Ogura,pert-CS,A note}. However, for superstrings we have the added complication that the CS theories are defined on Lie superalgebras, which is not a common feature of conventional CS theories\footnote{Superalgebra CS theories have been considered before in the literature albeit in a different context. See e.g. \cite{bra&sup,torsion}.}. The problem of quantizing our Hamiltonian double CS theory in the presence of the spectral parameter, i.e. the quantization of the Poisson bracket \eqref{pre Maillet}, is currently under investigation \cite{CS II} based on the combinatorial quantization approach of \cite{Fock-Rosly,combinatorial I,combinatorial II,Buffenoir-1,Buffenoir-2}. 


Several natural questions raise from these first steps and in what follows we mention some of them:

One potential application of this approach would be to study finite size effects. For $r \rightarrow \infty$, the boundary decompactifies and $\Sigma=S^{1}\times \mathbb{R}\rightarrow \mathbb{R}^{1,1}$. In this situation, the lambda model action must be carefully modified along the lines of \cite{WZW branes,NA kinks} in order to accommodate the new boundary conditions. In this limit though, asymptotic states and their S-matrix can be defined but for finite $r$ (to our knowledge) not much is known. It would be enlightening to study what the CS theory could tell us about the quantum integrability of the 1+1 theory for any value of $r$. A strategy for quantization would be to quantize the ultralocal theory on the disc and afterwards project the quantum theory to the boundary in some sort of holographic way (by imposing all the constraints). This is opposite to the usual symplectic reduction approach of first enforcing the constraints $F=0$ classically and then quantizing the remaining degrees of freedom.

The fundamental objects of our double CS theory would be the Wilson loops with spectral parameter
\begin{equation}
W_{R}(C;z)=\big\langle P\exp \big( -\oint_{C} \mathcal{A}(z)\big) \big\rangle _{R}, \label{Wilson}
\end{equation}
for $C$ a knot in $M=D\times \mathbb{R}$ and $R$ a particular representation. If $C$ is a horizontal curve constrained to $\partial D$ and wrapping it once, then we obtain\footnote{Other important objects would be the vertical $z$-dependent Wilson lines but at this moment their meaning is less clear.}
\begin{equation}
W_{R}(S^{1};z)=\big\langle P\exp \big(-\int\nolimits_{S^{1}} d\sigma \mathscr{L}^{\prime}(\sigma ;z)\big) \big\rangle _{R} \label{Mono}
\end{equation}
that is related to the monodromy of the extended Lax operator \eqref{extended Lax hybrid} and if we take $z=z_{\pm}$, then we get
\begin{equation}
W_{R}(S^{1};z_{\pm})=\big\langle P\exp \big(\pm \frac{2\pi }{k}\int\nolimits_{S^{1}}d\sigma 
\mathscr{J}_{\mp }(\sigma)\big) \big\rangle _{R} 
\end{equation}
that is related to the monodromy of the two Kac-Moody currents $\mathscr{J}_{\pm}$ as in \eqref{1 eq.}, leading to quantum groups \cite{hidden,quantum-group}. We should then expect a natural affine quantum group symmetry enhancement in our theory in terms of a quantum $R(z)$-matrix that (hopefully) is related to \eqref{R12} in the classical limit. Recently, in \cite{eta-affine} it is shown that this enhancement do occur for the eta models and this is done by expanding the monodromy matrix around the poles of the eta-deformed twisting function $\varphi_{\eta}(z)$ so it is reasonable to expect that this must be true for the lambda models as well as both theories are, in a sense, complementary. Another issue is related to the computation of the (classical/quantum) algebra of $z$-dependent Wilson loops \eqref{Wilson} defined on horizontal curves corresponding to the continuation of \eqref{Mono} into the interior of $D$. At this point it is too premature to make strong statements about its properties or even its existence but the results of \cite{CS-XXZ}, where a slightly similar situation is considered, suggest this algebra can be found\footnote{See \cite{PoissonStructure} an references therein as well as \cite{Rajeev-Turgut} for the standard $z$-independent case.} precisely by exploiting the lattice simulation approach of \cite{Fock-Rosly,combinatorial I,combinatorial II,Buffenoir-1,Buffenoir-2}. 

From \eqref{pre Maillet} and \eqref{R-bracket}, we realize that the Poisson structure is related to the symmetric operator $R+R^{*}$. This suggest a possible lift of the action \eqref{action on DxR} to the twisted loop superalgebra $\hat{\mathfrak{f}}$ in terms of the inner product \eqref{inner product on D}. The kinetic term in the Lagrangian \eqref{double CS Lag} should, in principle, be replaced by something of the form
\begin{equation}
L\sim \int\nolimits_{D}\left\langle A\wedge (R+R^{\ast })^{-1}\partial
_{\tau }A\right\rangle _{\varphi _{\lambda }}
\end{equation}
but we have not succeeded in showing it. In any case, it would be interesting to study if there is a connection of our CS formulation with the CS construction of lattice models presented in \cite{Witten,Costello-01,Costello-02}. In particular, if the action (2.8) of the paper \cite{Witten} could be related to our action \eqref{double action} for a 1-form $\omega \sim dz\varphi _{\lambda }(z)/z$.

A natural variation of our construction would be to investigate if the formulation in which the actions $S_{(+)}$ and $S_{(-)}$ are complex conjugated could be related to the eta models and if the results of \cite{Severa} could be applied to relate both formulations. 

The last question is how enters the lambda model dilaton field in the CS formulation after quantization is performed.

We will report on some of these questions elsewhere.

\section*{Acknowledgements}

The author would like thank T. J. Hollowood and J. L. Miramontes for valuable discussions and collaboration.

\end{document}